\documentclass{elsart}

\usepackage{amssymb}

\begin{document}

\begin{frontmatter}



\title{Arbitrary precision composite pulses for NMR quantum computing}


\author{William G. Alway}, \author{Jonathan A. Jones\corauthref{cor}}\ead{jonathan.jones@qubit.org}
\corauth[cor]{Corresponding author.}
\address{Oxford Centre for Quantum Computation, Clarendon Laboratory, Parks Road, Oxford OX1~3PU, UK}

\begin{abstract}
We discuss the implementation of arbitrary precision composite
pulses developed using the methods of Brown \textit{et al.} [Phys.
Rev. A 70 (2004) 052318].  We give explicit results for pulse
sequences designed to tackle both the simple case of pulse length
errors and for the more complex case of off-resonance errors. The
results are developed in the context of NMR quantum computation, but
could be applied more widely.
\end{abstract}

\end{frontmatter}

\section{Introduction}
Composite pulses \cite{levitt86} have long played an important role
in many NMR experiments, allowing the effects of systematic errors
to be reduced.  More recently they have been applied in quantum
computing \cite{bennett00}, including both magnetic resonance
experiments and other implementations \cite{cummins00, cummins03,
collin04, morton05b}. Composite pulses developed for quantum
computing differ from more conventional NMR approaches in two key
ways. Firstly they must perform the desired rotation whatever the
starting state of the system, so that they act as \textit{general
rotors}; in NMR such pulses are sometimes called type~A composite
pulses \cite{levitt86}, and have the advantage that they can be
inserted into any part of a pulse sequence without the need for
careful analysis. Secondly, they are usually designed to give
extremely accurate rotations in the presence of small errors, rather
than to give reasonable rotations in the presence of large errors.
Despite these differences, however, composite pulses designed for
quantum computation can be used in conventional NMR experiments, and
studying them can give some insight into more conventional
approaches.

Type~A composite pulses are most commonly used to tackle pulse
length errors, which occur when the strength of the driving field
differs from its nominal value, for example as a result of
inhomogeneity, and off-resonance errors, which occur when the
frequency of the driving field is not quite in resonance with the
transition of interest. For pulse length errors the BB1 sequence
\cite{wimperis94}, originally developed by Wimperis, has proved
highly successful.  Designing good type~A pulses for off-resonance
errors has proved more difficult, with the CORPSE sequence
\cite{cummins00, cummins03}, based on an early numeric result by
Tycko \cite{tycko83}, being perhaps the most satisfactory.

More recently Brown \textit{et al.} have described a general method
\cite{brown04, brown05} for generating \textit{arbitrary precision}
composite pulses, which they claim can be applied to tackle errors
of arbitrary kinds. They do not, however, give explicit solutions
for more than a small number of cases.  We will show that it is
simple to apply their methods to generate pulses resistant to pulse
length errors, but it is more difficult to tackle off-resonance
effects.

\section{Pulse length errors}
We consider rotations about axes in the $xy$ plane, for which an
ideal rotation is described by the propagator
\begin{equation}
U(\theta,\phi)=\exp[-i\,\theta(\sigma_x\cos\phi+\sigma_y\sin\phi)/2]
\end{equation}
where $\theta$, the rotation angle, depends on the strength of the
resonant driving field and the time for which it is applied, and
$\phi$, the phase, depends on the phase of the field with respect to
some appropriate reference.  Following Brown \textit{et al.}
\cite{brown04} we describe our rotations in terms of the Pauli
matrices, which are trivially related to the corresponding single
spin product operator terms \cite{sorensen89}. The description above
is really only appropriate when $\theta\ge0$, but also works for
negative angles as
\begin{equation}
U(-\theta,\phi)=U(\theta,\phi+\pi)=U^{-1}(\theta,\phi).
\end{equation}
In the presence of errors, however, this extension may not be
appropriate and it is necessary to proceed with some caution.

Pulse length errors occur when the strength of the field is higher
or lower than its nominal value, so that all rotation angles are
systematically wrong by some constant fraction $\epsilon$.  It is
convenient to write
\begin{equation}
V(\theta,\phi)=U(\theta[1+\epsilon],\phi)
\end{equation}
in such cases, and a series expansion in the error $\epsilon$ gives
\begin{equation}
V(\theta,\phi)=U(\theta,\phi)+O(\epsilon)
\end{equation}
so that the propagator has first order errors in $\epsilon$.  Note
that in this paper we write all our composite pulses as sequences of
propagators, so that the order of pulses runs from right to left. We
classify pulse sequences according to $n$, the order of error in the
propagator, but will occasionally refer to the corresponding
propagator fidelity
\begin{equation}
\mathcal{F}=\frac{1}{2}\textrm{Tr}(VU^{-1})=1-O(\epsilon^{2})
\end{equation}
in which errors appear to order $2n$.  For pulse length errors
\begin{equation}
V(\theta,\phi+\pi)=V^{-1}(\theta,\phi)
\end{equation}
which will prove very useful throughout this section.

Several composite pulse methods for correcting such pulse length
errors exist, most notably BB1 \cite{wimperis94}. This replaces a
single pulse with a sequence of pulses such that
\begin{equation}
V(\pi,\phi_a)V(2\pi,\phi_b)V(\pi,\phi_a)V(\theta,0)=U(\theta,0)+O(\epsilon^3)
\end{equation}
where $\phi_b=3\phi_a$ and $\phi_a=\arccos(-\theta/4\pi)$, to give a
propagator with third order errors in $\epsilon$.  This pulse
sequence was discovered using geometric arguments, and turns out to
be remarkably effective in practice \cite{cummins03, xiao06}.

Brown \textit{et al.} replaced previous methods of finding composite
pulses, based on intuitions or special forms, by a systematic
procedure \cite{brown04}.  We begin by describing in detail how this
procedure can be used to generate a series of composite pulses to
correct pulse length errors.  As before we only consider target
rotations with phase angles of zero, as more general rotations in
the $xy$ plane are trivially derivable from these by offsetting the
phase of all pulses appropriately.  These pulses can also be used to
design sequences for robust evolution under J-couplings
\cite{jones03}.

\subsection{Isolating the error}
The first step is to isolate the error part of a pulse sequence by
calculating
\begin{equation}
A_1=V(\theta,0)\,U^{-1}(\theta,0)= \mathbf{1}
-\frac{i\theta}{2}\,\epsilon\,\sigma_x +O(\epsilon^2)
\end{equation}
where $A_1$ indicates a first-order error in a sequence correct to
zero-order. Since
\begin{equation}
A_1^{-1}V=UV^{-1}V=U
\end{equation}
if we can generate $A_1^{-1}$
\textit{exactly} then we can convert an erroneous rotation into a
correct one.  More realistically, if we can generate $A_1^{-1}$
\textit{correct to first order} then we can use this to cancel out
the first order error in the original pulse. To do this it is useful
to note that
\begin{equation}
A_1=U(\epsilon\theta,0) + O(\epsilon^2)
\end{equation}
so that $A_1^{-1}$ is approximately equal to an $x$-rotation with
angle $-\theta\epsilon$.

\subsection{Generating a pure error term}
To correct the first-order error it is necessary to generate a
matching rotation whose angle is \textit{directly proportional} to
the error fraction $\epsilon$.  A pure error term of this kind is
most easily generated by noting that
\begin{equation}
V(2\pi,0)= -\mathbf{1} +i\pi\epsilon\,\sigma_x +O(\epsilon^2)
\end{equation}
has the correct general form (the global phase of $-1$ can be
ignored as usual).  However while the error has the right form it
has the wrong magnitude.  This can be fixed by using two rotations
with different phases \cite{brown04}, as
\begin{equation}
X_1(\phi)=V(2\pi,\phi)V(2\pi,-\phi)=\mathbf{1}-i\,2\pi\cos(\phi)\,\epsilon\,\sigma_x
+O(\epsilon^2) \label{eq:x1}
\end{equation}
and so the size of the error can be scaled.  In particular, solving
\begin{equation}
-2\pi\cos(\phi_1)=\theta/2
\end{equation}
to get
\begin{equation}
\phi_1=\pm\arccos(-\theta/4\pi)\label{eq:phi1}
\end{equation}
allows the first order error to be corrected (note that the sign of
the error term must be reversed as we are seeking to approximate
$A_1^{-1}$ rather than $A_1$). This gives
\begin{equation}
X_1(\phi_1)V(\theta,0)=U(\theta,0)+O(\epsilon^2)
\end{equation}
as a pulse sequence which is correct to second order. Interestingly
the key phase angle $\phi_1$ has the same size as $\phi_a$ in the
BB1 sequence. The sign of $\phi_1$ (and $\phi_a$) seems arbitrary,
but for definiteness we choose the positive value.  This point will
be examined in more detail later.

The size and scaling of the errors permits a value of $\phi_1$ to be
found for any choice of $\theta$, but if larger error terms are
needed they can be achieved by simply repeating the sequence of
$2\pi$ rotations, thus doubling the error. In passing we note that
although the approximate form given in Eq.~\ref{eq:x1} is
independent of the sign of $\phi$, and thus of the order of the two
$2\pi$ pulses, the exact form of $X_1$ does in fact depend on the
sign of $\phi$, and it is therefore necessary to use a consistent
order.  This will become important below.

\subsection{Treating the second order error}
The process described above can then be repeated to isolate the
second order error
\begin{equation}
A_2=X_1(\phi_1)V(\theta,0)\,U^{-1}(\theta,0)
\end{equation}
with the result
\begin{equation}
A_2=\mathbf{1}-\frac{i\theta\sqrt{16\pi^2-\theta^2}}{8}\,\epsilon^2\,\sigma_z+O(\epsilon^3).
\end{equation}
To correct this will require an error term dependent on
$\epsilon^2$, rather than on $\epsilon$, and directed along the
$z$-axis rather than the $x$-axis.  This can be achieved by using
the properties of \textit{group commutators}.  As these properties
play a key role in the proof of the Solovay--Kitaev theorem
\cite{dawson06} Brown \textit{et al.} refer to the results as SK
pulse sequences, but beyond the role of group commutators there is
no need to undertand the Solovay--Kitaev theorem to see how their
methods work. The key result
\begin{eqnarray}
\exp(-iA\epsilon^l)\exp(-iB\epsilon^m)\exp(iA\epsilon^l)\exp(iB\epsilon^m)\nonumber\\*
=\exp([A,B]\epsilon^{l+m})+O(\epsilon^{l+m+1})
\end{eqnarray}
shows that two different first order pure error terms can be
combined to make a single second order pure error term as long as
their directions are chosen properly \textit{and} inverses are
available for each error term.  This restriction will not be
important for pulse length errors but will be more problematic for
off-resonance errors.

As the commutator $[\sigma_x,\sigma_y]=-2i\sigma_z$ the desired
second order error term along $z$ can be generated from $x$ and $y$
terms.  The $x$ term can be generated as before, Eq.~\ref{eq:x1},
and the equivalent $y$ term $Y_1$ can be generated in the same way
by shifting the phase of both pulses by $\pi/2$.  (Brown \textit{et
al.} in fact described an alternative method \cite{brown04} for
implementing $Y_1$, but this alternative is more complex.) Inverse
terms can be generated by reversing the order of the two $2\pi$
pulses and shifting their phases by $\pi$. Note that it is necessary
to reverse the sequence of pulses even though Eq.~\ref{eq:x1}
appears to be independent of this, as the two alternatives differ to
second order in $\epsilon$; it is \textit{not} sufficient to
generate an inverse which only accurate to first order.  Hence
\begin{equation}
Z_2(\phi)=X_1(\phi)Y_1(\phi)X_1^{-1}(\phi)Y_1^{-1}(\phi)=\mathbf{1}-i\,8\pi^2\cos^2(\phi)\,\epsilon^2\,\sigma_z+O(\epsilon^3)\label{eq:z2}
\end{equation}
allows a second order $z$ error of some desired size to be
generated.  If an error of the opposite sign is needed then the
$X_1$ and $Y_1$ sequences can be interchanged to give
\begin{equation}
Z_2'(\phi)=Y_1(\phi)X_1(\phi)Y_1^{-1}(\phi)X_1^{-1}(\phi)=\mathbf{1}+i\,8\pi^2\cos^2(\phi)\,\epsilon^2\,\sigma_z+O(\epsilon^3).
\end{equation}
As before we can solve
\begin{equation}
8\pi^2\cos^2(\phi_2)=\theta\sqrt{16\pi^2-\theta^2}/8
\end{equation}
to remove the second order error, giving
\begin{equation}
\phi_2=\pm\arccos\left[\pm\frac{\sqrt[4]{16\pi^2\theta^2-\theta^4}}{8\pi}\right]\label{eq:phi2}
\end{equation}
where the signs may be chosen independently, and we choose initially
to take both positive signs.  Thus
\begin{equation}
Z_2'(\phi_2)X_1(\phi_1)V(\theta,0)=U(\theta,0)+O(\epsilon^3)
\label{eq:sk2}
\end{equation}
is our desired pulse sequence, correct to third order.  The explicit
expansion
\begin{eqnarray}
V(2\pi,\pi/2+\phi_2)V(2\pi,\pi/2-\phi_2)V(2\pi,\phi_2)V(2\pi,-\phi2)\nonumber\\*
V(2\pi,3\pi/2-\phi_2)V(2\pi,3\pi/2+\phi_2)V(2\pi,\pi-\phi2)V(2\pi,\pi+\phi_2)\nonumber\\*
V(2\pi,\phi_1)V(2\pi,-\phi1)V(\theta,0)
\end{eqnarray}
shows that this contains a total of ten correction pulses in
addition to the main pulse.

\subsection{Rotating and redividing the error}
The sequence described above is not in fact the sequence originally
described by Brown \textit{et al.}  Their approach \cite{brown04} is
instead based on generating an $X_2$ error correction term using
appropriate $Y_1$ and $Z_1$ sequences and then rotating this onto
the $z$ axis.  In the absence of systematic errors, such rotations
are easily performed.  For example the identity
\begin{equation}
U(\pi/2,3\pi/2)U(\theta,0)U(\pi/2,\pi/2)=\exp(-i\theta\,\sigma_z/2)
\end{equation}
allows a $z$ rotation to be generated from $x$ and $y$ rotations, a
composite Z-pulse \cite{freeman81}.  In the presence of systematic
errors it is necessary to proceed with more caution, as the errors
in different pulses will combine in a complex manner.  However, in
the presence of pulse length errors it is possible to use imperfect
pulses to rotate pure error terms, as
\begin{equation}
V(\pi/2,3\pi/2)X_nV(\pi/2,\pi/2)=Z_n+O(\epsilon^{n+1})
\end{equation}
for any error order $n$, and similarly for other error terms. This
approach requires the ability to generate accurate inverse
operators, and while this is easy for pulse length errors it can be
tricky in other cases.

For pulse-length errors we can implement the second order correction
sequence using
\begin{eqnarray}
Z_2'(\phi_2)&\approx&V(\pi/2,3\pi/2)X_2'(\phi_2)V(\pi/2,\pi/2)\label{eq:z2a}\\*
X_2'(\phi_2)&=&Z_1(\phi_1)Y_1(\phi_1)Z_1^{-1}(\phi_1)Y_1^{-1}(\phi_1)\\*
Z_1(\phi_2)&\approx&V(\pi/2,3\pi/2)X_1(\phi_2)V(\pi/2,\pi/2)\label{eq:z2c}.
\end{eqnarray}
This alternative sequence gives identical performance at second
order (complete correction of errors) but differs in its third order
behaviour, as we shall see later.

We can also consider many other possibilities: firstly we can
instead generate $Z_2'$ from any of $Y_2'$, $X_2$, or $X_2'$;
secondly we can use alternative rotations to generate $Z_1$; thirdly
we can use the negative sign for $\phi_1$ in the first order
correction sequence (in which case the second order error changes
sign). Beyond these possibilities, built around rotating the error,
we can also choose how to divide up the relative contributions to
the second order error term arising from the two first order terms.
For example we can write
\begin{eqnarray}
Z_2(\alpha,\beta)&=&X_1(\alpha)Y_1(\beta)X_1^{-1}(\alpha)Y_1^{-1}(\beta)\nonumber\\*
&=&\mathbf{1}-i\,8\pi^2\cos(\alpha)\cos(\beta)\,\epsilon^2\,\sigma_z+O(\epsilon^3)
\end{eqnarray}
and control the size of the second order error term by varying
$\alpha$ and $\beta$.  More simply still, we can use the fact that
$V(2\pi,0)$ gives an \textit{unscaled} pure error term along $x$ to
use the form
\begin{equation}
V(2\pi,0)Y_1(\beta)V(2\pi,\pi)Y_1^{-1}(\beta)=\mathbf{1}-i\,4\pi^2\cos\beta\,\epsilon^2\,\sigma_z+O(\epsilon^3)
\end{equation}
which only has six pulses, rather than the usual eight, and obtain
the correct error term by choosing
\begin{equation}
\beta=\pm\arccos\left[-\frac{\theta\sqrt{16\pi^2-\theta^2}}{32\pi^2}\right].
\end{equation}

\subsection{Choosing between sequences}
Given this plethora of subtly different sequences it is reasonable
to ask which is the best.  In some sense \textit{all} second-order
pulse sequences are equally good, as they all suppress errors to the
same order, but it is possible to choose between them either by
considering higher order errors, or by considering sensitivity to
\textit{other} types of error \cite{cummins03}.

Here we adopt the first approach, choosing to minimize the size of
the third order error term, and initially specializing to the case
of $180^\circ$ pulses, so that $\theta=\pi$.  The smallest error
term we have so far been able to find occurs when using equations
\ref{eq:z2a} to \ref{eq:z2c} to create the second order term, and
taking positive signs throughout equations \ref{eq:phi1} and
\ref{eq:phi2}.  There is no obvious reason \textit{why} this choice
is best, but it does give a third order error more than 15 times
smaller than some other alternatives.

Interestingly, the second best choice we have found is the BB1
sequence, which for the case $\theta=\pi$ has an error only about
10\% larger than the best sequence.  Furthermore for BB1 the size of
the third order error term scales approximately linearly with
$\theta$, while the behavior of the ``best'' sequence, described
above, is more complex.  Thus for most flip angles (specifically,
$\theta<168^\circ$) BB1 is the best second order sequence known. For
the case of $180^\circ$ pulses it has a fidelity
\begin{equation}
\mathcal{F}=1-\frac{5\pi^6}{1024}\,\epsilon^6+O(\epsilon^8).
\end{equation}

\subsection{Third order errors}
We can of course correct the third order error term in much the same
way as the first and second order errors.  As pointed out by Brown
\textit{et al.} there is no need to use a fully systematic approach
of correcting error orders in sequence; instead we can begin with
BB1 and correct the third order error term.

The third order error term for BB1 (and indeed for all the other
sequences considered above) lies in the $xy$ plane, with the size
and position depending on the value of $\theta$.  Here we do not
give complete results, but simply sketch a partial solution.  There
are many possibilities for generating a third order error, but one
simple example is
\begin{equation}
X_3(\phi)=Y_1(\phi)Z_2(\phi)Y_1^{-1}(\phi)Z_2^{-1}(\phi)=\mathbf{1}-i\,32\pi^3\cos^3(\phi)\,\epsilon^3\,\sigma_x+O(\epsilon^4)
\end{equation}
where $Z_2(\phi)$ is generated from $X_1(\phi)$ and $Y_1(\phi)$ as
before, equation \ref{eq:z2}, and $Z_2^{-1}=Z_2'$, so that this
sequence only requires $x$ and $y$ rotations. The error term can
then be rotated into the correct position, either by composite $z$
rotations or, more simply, by shifting the phases of all the pulses
in the $X_3$ term.

As before we need to choose a value of $\phi_3$ to cancel the third
order error, but as the calculations become very complicated we here
consider only the special case of $180^\circ$ pulses, $\theta=\pi$.
Even in this case the analytic result is complicated, and so we
simply give the numerical value, $\phi_3\approx73.1^\circ$; the
required phase shift can also be calculated as approximately
$-1.6^\circ$.

\section{Off-resonance errors}
The treatment of off-resonance errors is superficially similar but
much more difficult in practice.  The fundamental problem is that,
unlike the case of pulse length errors, it is not possible to
generate perfect inverses of arbitrary rotations in the presence of
off-resonance errors.

Off-resonance errors occur when the frequency of the driving field
is not quite in resonance with the transition of interest, so that
all rotations occur around a tilted axis. They can be parameterized
in terms of the off-resonance fraction $f$, equal to the ratio of
the frequency error and the driving frequency.  The basic rotation
in then
\begin{equation}
V(\theta,\phi)=\exp[-i\,\theta(\sigma_x\cos\phi+\sigma_y\sin\phi+f\,\sigma_z)/2]=U(\theta,\phi)+O(f)
\end{equation}
which has first order errors in $f$. Unlike the case of pulse-length
errors this definition should \textit{only} be used for positive
values of $\theta$.

The CORPSE family of sequences \cite{cummins00, cummins03} for
correcting off-resonance errors uses the three pulse sequence
\begin{equation}
C(\theta,\phi)=V(\theta_c,\phi)V(\theta_b,\phi+\pi)V(\theta_a,\phi)=U(\theta,\phi)+O(f^2)
\end{equation}
where
\begin{eqnarray}
\theta_a&=&n_a\,2\pi+\theta/2-\arcsin[\sin(\theta/2)/2]\\*
\theta_b&=&n_b\,2\pi-2\arcsin[\sin(\theta/2)/2]\\*
\theta_c&=&n_c\,2\pi+\theta/2-\arcsin[\sin(\theta/2)/2]
\end{eqnarray}
and the size of the second order error term depends on the values
chosen for the integers $n_a$, $n_b$ and $n_c$.  The smallest errors
are seen for the original CORPSE sequence, which has $n_a=n_b=1$ and
$n_c=0$; for the case of $180^\circ$ pulses the fidelity is
\begin{equation}
\mathcal{F}=1-\frac{12+\pi^2-4\sqrt{3}}{32}\,f^4+O(f^6).
\end{equation}
Short-CORPSE, defined by $n_a=n_c=0$ and $n_b=1$ is the shortest
possible sequence but has a much larger error term \cite{cummins03}.

The CORPSE family will play a key role in the following sections,
not simply because it provides a pulse sequence with no first order
errors, but mostly because it provides a route to sufficiently
accurate inverse propagators.  In the presence of off-resonance
errors $V(\theta,\phi+\pi)$ is \textit{not} an accurate inverse for
$V(\theta,\phi)$ as
\begin{equation}
V(\theta,\pi)V(\theta,0)=\mathbf{1}+O(f).
\end{equation}
The corresponding CORPSE pulses perform much better,
\begin{equation}
C(\theta,\pi)C(\theta,0)=\mathbf{1}+O(f^3),
\end{equation}
and provide sufficiently accurate inverses to allow error terms to
be rotated as for pulse length errors.  As before the size of the
third order error term depends on the exact choice of sequence, but
is now smallest for short-CORPSE.

\subsection{Correcting first order errors}
We now explore the systematic correction of error orders using the
methods previously described.  The first order error can be isolated
as before,
\begin{eqnarray}
A_1&=&V(\theta,0)U^{-1}(\theta,0)\nonumber\\*
&=&\mathbf{1}-i\,\sin(\theta)/2\,f\,\sigma_z+i\,\sin^2(\theta/2)\,f\,\sigma_y+O(f^2)
\end{eqnarray}
and in general lies in the $yz$ plane. Tunable pure error terms can
be created either by using the form given by Brown \textit{et al.}
\cite{brown04},
\begin{equation}
B_1(\phi)=V(\phi,0)V(2\phi,\pi)V(\phi,0)=\mathbf{1}-i\,2\sin(\phi)\,f\,\sigma_z+O(f^2),
\end{equation}
or the alternative form
\begin{eqnarray}
Y_1'(\phi)&=&V(\pi,\phi)V(\pi,\pi+\phi)V(\pi,-\phi)V(\pi,\pi-\phi)\nonumber\\*
&=&\mathbf{1}+i\,4\cos(\phi)\,f\,\sigma_y+O(f^2).
\end{eqnarray}

Designing a sequence for the case $\theta=\pi$ is easy as the error
lies solely along the $y$ axis in this case, and so
\begin{equation}
Y_1'(\phi_1)V(\pi,0)=U(\pi,0)+O(f^2)\label{eq:or1}
\end{equation}
withe the choice $\phi_1=\arccos(-1/4)\approx104.5^\circ$.
Interestingly, the key phase angle in this sequence turns out to be
the same as that used in a BB1 pulse with $\theta=\pi$.  The size of
the second order error term is significantly larger than for CORPSE
(and somewhat larger than for short-CORPSE), with a fidelity
\begin{equation}
\mathcal{F}=1-\frac{60+\pi^2}{32}\,f^4+O(f^5).
\end{equation}
but this sequence does have the relative simplicity of being
constructed solely from $180^\circ$ rotations, albeit with
complicated phases.

Designing a sequence for other values of $\theta$ is, however, much
trickier.  In the general case the error does not lie along a
principal axis, and so it might seem that we should rotate one of
the two pure error terms. This cannot be done directly using simple
rotations, as accurate inverse propagators are required to rotate
error terms. It could be achieved using CORPSE pulses, but this is
not sensible as CORPSE is already correct to first order.

An alternative approach is to note that pure error sequences can
simply be combined, and so build up a tilted error by combining the
$z$ and $y$ error sequences,
\begin{equation}
B_1(\phi_1^z)Y_1'(\phi_1^y)=\mathbf{1}-i\,2\sin(\phi_1^z)\,f\,\sigma_z+i\,4\cos(\phi_1^y)\,f\,\sigma_y+O(f^2)
\end{equation}
with any cross terms between the two parts if the pulse sequence
being swallowed up into the $O(f^2)$ term. Choosing
$\phi_1^y=\arccos[-\sin^2(\theta/2)/4]$ and
$\phi_1^z=-\arcsin[\sin(\theta)/4]$ allows first order off-resonance
errors to be suppressed in the general case.
However, the size of the second order error term remains
significantly larger than for CORPSE, and this general sequence does
not have the simplicity seen for the special case of $180^\circ$
pulses. Thus CORPSE remains the best currently known type~A sequence
correct to first order in the presence of off-resonance errors.

\subsection{Correcting higher order errors}
The systematic approach can, more sensibly, be used to correct
higher order errors.  As the equations for arbitrary values of
$\theta$ become extremely complicated we will again limit ourselves
to the special case $\theta=\pi$.  We extend our definitions
\begin{eqnarray}
X_1(\phi)&=&V(\pi,\pi/2-\phi)V(\pi,3\pi/2-\phi)V(\pi,\pi/2+\phi)V(\pi,3\pi/2+\phi)\nonumber\\*
&=&\mathbf{1}-i\,4\cos(\phi)\,f\,\sigma_x+O(f^2)\\*
Y_1(\phi)&=&V(\pi,\pi-\phi)V(\pi,-\phi)V(\pi,\pi+\phi)V(\pi,\phi)\nonumber\\*
&=&\mathbf{1}-i\,4\cos(\phi)\,f\,\sigma_y+O(f^2)\\*
X_1'(\phi)&=&V(\pi,3\pi/2+\phi)V(\pi,\pi/2+\phi)V(\pi,3\pi/2-\phi)V(\pi,\pi/2-\phi)\nonumber\\*
&=&\mathbf{1}+i\,4\cos(\phi)\,f\,\sigma_x+O(f^2)\\*
\end{eqnarray}
and note that $X_1'(\phi)\approx X_1^{-1}(\phi)$.  The approximation
is good enough that these terms can be used to build higher order
propagators; in particular
\begin{equation}
Z_2(\phi)=X_1(\phi)Y_1(\phi)X_1'(\phi)Y_1'(\phi)=\mathbf{1}-i\,32\cos^2(\phi)\,f^2\,\sigma_z
\end{equation}
and
\begin{equation}
Z_2'(\phi)=Y_1(\phi)X_1(\phi)Y_1'(\phi)X_1'(\phi)=\mathbf{1}+i\,32\cos^2(\phi)\,f^2\,\sigma_z.
\end{equation}
The second order error can be isolated as usual
\begin{eqnarray}
A_2&=&Y_1'(\phi_1)V(\pi,0)U^{-1}(\pi,0)\nonumber\\*
&=&\mathbf{1}-i\,\sqrt{15}/2\,f^2\,\sigma_z-i\,\pi/4\,f^2\,\sigma_x+O(f^3)
\end{eqnarray}
and lies in the $xz$ plane with magnitude $\sqrt{60+\pi^2}/4$.

One approach to correcting this is by using CORPSE pulses to rotate
an appropriate $Z_2$ error around the $y$ axis to get the final
sequence
\begin{equation}
C(\psi_2,\pi/2)Z_2'(\phi_2)C(\psi_2,3\pi/2)Y_1'(\phi_1)V(\pi,0)=U(\pi,0)+O(f^3)
\end{equation}
with
\begin{equation}
\phi_2=\arccos\left(\frac{\sqrt[4]{60+\pi^2}}{8\sqrt{2}}\right)\approx75.2^\circ
\end{equation}
and
\begin{equation}
\psi_2=\arctan\left(\frac{\pi}{2\sqrt{15}}\right)\approx22.1^\circ.
\end{equation}
This approach can, of course, be generalized to other angles and
higher orders, but the resulting algebra is very complex.  For
simplicity it is possible to check results using ideal rotations in
place of CORPSE based rotations: this will give the right result for
error terms which are completely suppressed, but the wrong values
for higher order errors.

Alternatively, we can construct the tilted error term out of a
combination of $z$ and $x$ errors as in the previous section.  We
begin by noting that $B_1(3\pi/2)$ provides a good inverse for the
pure error term $B_1(\pi/2)$, and that this allows us to construct a
second order $x$ error using
\begin{equation}
X_2(\phi)=Y_1(\phi)B_1(\pi/2)Y_1'(\phi)B_1(3\pi/2)=\mathbf{1}-i\,16\cos(\phi)\,f^2\,\sigma_x.
\end{equation}
This can be combined with a $z$ error to get the pulse sequence
\begin{equation}
X_2(\phi_2^x)Z_2'(\phi_2^z)Y_1'(\phi_1)V(\pi,0)=U(\pi,0)+O(f^3)
\end{equation}
where $\phi_2^x=\arccos(-\pi/64)\approx92.8^\circ$ and
$\phi_2^z=\arccos(\sqrt[4]{15}/8)\approx75.8^\circ$.  This sequence
has the advantage over the CORPSE based approach of requiring only
$90^\circ$ and $180^\circ$ pulses.

\subsection{Time symmetric sequences}
In passing we consider the use of time-symmetry in composite pulse
sequences \cite{cummins03}.  In the presence of off-resonance errors
time symmetric composite pulses have fidelities which are even
functions of the off-resonance fraction $f$ \cite{bowdrey03}, and so
give the same fidelity for $+f$ and $-f$, although the details of
the error may differ.  Although such symmetric fidelities have no
advantage in principle, the results are certainly easier to
interpret.

As an example we consider a time symmetric version of the pulse
sequence to correct first order errors arising from off-resonance
effects in a $180^\circ$ pulse, which takes the form
\begin{equation}
Y_1'(\phi_1')V(\pi,0)Y_1(\phi_1')=U(\pi,0)+O(f^2)
\end{equation}
where $\phi_1'=\arccos(-1/8)\approx97.2^\circ$.  The fidelity of
this sequence has no fifth order term, unlike that of the previous
version, equation \ref{eq:or1}, but as both fidelities are dominated
by fourth order errors this is largely a cosmetic improvement.

\subsection{Simultaneous errors} So far we have only considered the
effects of pulse length errors and off-resonance errors in
isolation, while in real physical systems both sorts of error can
occur simultaneously. It is generally difficult to find pulse
sequences which suppress both sorts of error simultaneously, but it
is still important to consider whether insensitivity to one type of
error is obtained at the cost of increased sensitivity to other
types of errors \cite{cummins03}.

As noted previously \cite{cummins03}, the response of the time
symmetric version of BB1 to off-resonance errors is very similar to
that of a simple pulse.  This occurs because \textit{in the absence
of pulse length errors} the correction sequence
\begin{equation}
V(\pi,\phi_1)V(2\pi,3\phi_1)V(\pi,\phi_1)=\mathbf{1}+O(f^2)
\end{equation}
has no first order terms arising from off-resonance errors, and so
does not contribute significantly to the total error.  In the same
way \textit{in the absence of off-resonance errors} the correction
sequence
\begin{equation}
V(\pi,\phi_1)V(\pi,\pi+\phi_1)V(\pi,-\phi_1)V(\pi,\pi-\phi_1)=\mathbf{1}
\end{equation}
has no error terms arising from pulse length errors at all.  Thus
these two sequences can be combined, giving the composite
$180^\circ$ pulse
\begin{eqnarray}
V(\pi,\phi_1)V(\pi,\pi+\phi_1)V(\pi,-\phi_1)V(\pi,\pi-\phi_1)\nonumber\\* V(\pi,\phi_1)V(2\pi,3\phi_1)V(\pi,\phi_1)V(\pi,0)
\end{eqnarray}
with $\phi_1=\arccos(-1/4)$.  This has a fidelity
\begin{equation}
\mathcal{F}=1-\frac{15}{8}\,f^4-\frac{5\pi^6}{1024}\,\epsilon^6-\frac{169\pi^2}{32}\,f^2\epsilon^2+\textit{higher
terms}
\end{equation}
so in the absence of off-resonance errors the correction of pulse
length errors is identical to a BB1 sequence, and in the absence of
pulse length errors the correction of off-resonance errors is even
better than the simple pulse, equation~\ref{eq:or1}.  This pulse can
correct well for \textit{either} pulse length errors \textit{or}
off-resonance errors; in the presence of \textit{simultaneous}
errors the performance is not so good, but is still much better than
a simple $180^\circ$ pulse.

\section{Summary}
The methods of Brown \textit{et al.} can indeed be used to derive
arbitrary precision composite pulses, but the process can be
somewhat complicated.  For the case of pulse length errors the
situation is simple, as it is possible both to generate a wide range
of pure error terms and their inverses, and to rotate these terms
using uncorrected pulses.  The case of off-resonance errors is much
more complicated because the difficulty of generating accurate
inverses of incorrect rotations means that the most direct approach
cannot be used.

\section*{Acknowledgements}
We thank Mark Jerzykowski and Kenneth Brown for helpful
conversations, and the UK EPSRC and BBSRC for financial support.



\end{document}